\documentclass[preprint,showpacs,showkeys,floatfix,preprintnumbers,amsmath,amssymb]{revtex4}

\usepackage{graphicx}% Include figure files
\usepackage{verbatim}
\usepackage{color}

\begin{document}

\bibliographystyle{apsrev}

\title{Local quasiparticle density of states of
    superconducting SmFeAsO$1-x$F$x$ single crystals: Evidence for
    spin-mediated pairing }

\author{Y. Fasano$1,$\cite{Bariloche}, I. Maggio-Aprile$1$,
N.D. Zhigadlo$2$, S. Katrych $2$, J. Karpinski$2$ and {\O}.
Fischer$1$}

\affiliation{$1$  DPMC-MaNEP, University of Geneva, Switzerland}

\affiliation{$2$ Laboratory for Solid State Physics, ETH Zurich,
Switzerland}

\date{\today}

\begin{abstract}

We probe the local quasiparticles density-of-states in
micron-sized SmFeAsO$_{1-x}$F$_{x}$ single-crystals by means of
Scanning Tunnelling Spectroscopy. Spectral features resemble those
of cuprates, particularly a dip-hump-like structure developed at
energies larger than the gap that can be ascribed to the coupling
of quasiparticles to a collective mode, quite likely a resonant
spin mode.  The energy of the collective mode revealed in our
study decreases when the pairing strength increases. Our findings
support spin-fluctuation-mediated pairing in pnictides.

\end{abstract}

\pacs{74.50.+r, 74.20.Mn, 74.72.Hs}

\keywords{pnictide superconductors, pairing mechanism, scanning
tunnelling spectroscopy}

\maketitle

The recent discovery of superconductivity in oxypnictides renewed
the hopes on  solving the longstanding puzzle of the pairing
mechanism in high-temperature superconductors (HTS).
 This expectation is based on pnictides sharing resemblances with
cuprate-HTS: superconductivity emerges when charge-doping the
magnetic parent material \cite{delaCruz-2008a}, and $T_{c}$ can be
tuned with dopant concentration \cite{Luetkens-2008a}.
Spin-fluctuation-mediated pairing has been proposed as a promising
candidate for explaining high-temperature superconductivity in
oxypnictides \cite{Mazin-2008a,Wang-2009a,Kuroki-2008a}. In this
context, electron-tunneling data can provide crucial evidence: the
quantitative explanation of particular features in the
electron-tunneling spectra of conventional superconductors led to
unveiling their pairing mechanism
\cite{Giaever-1962,McMillan-1965}. We use scanning tunneling
microscopy (STM)  to probe the local  quasiparticle excitation
spectrum of  single-crystals of SmFeAsO$_{1 - x}$F$_{x}$
\cite{Zhigadlo-2008a,Karpinski-2009a}.
 We detect spectral features that fingerprint the coupling of
quasiparticles with a collective excitation, quite likely a
 spin resonant-mode \cite{Christianson-2008a,Wakimoto-2009a}. We find that the energy
of the STM-revealed collective mode decreases when the local
pairing strength increases, in remarkable similarity with the
phenomenology observed in cuprates \cite{Eschrig-2006}.

Among the  similarities between pnictide and cuprate-HTS, their
layered crystal structure is fundamental to compare novel
electron-tunneling data in pnictides with exhaustive studies in
cuprates \cite{Fischer-2007a}.
In iron-pnictides, superconductivity
is obtained by adding holes or electrons to the FePn layers
(Pn: pnictogen atom)  of the undoped parent compound.
Depending on the number of FeAs planes per crystal unit
cell (one or two), 1111 and 122-pnictides can be synthesized
\cite{Karpinski-2009a}. As in cuprates, the electronic structure
of pnictides is quasi-two-dimensional \cite{Singh-2008a}.
Increasing the list of resemblances, a resonant magnetic
excitation localized in energy and wave-vector has been detected
within the superconducting phase of 122 \cite{Christianson-2008a}
and very recently in 1111-pnictides \cite{Wakimoto-2009a}. These last two
 findings suggest that information on the pnictides' pairing mechanism
 can be obtained by analyzing electron-tunneling data with a
model \cite{Eschrig-2000} that considers the coupling of
quasiparticles with a collective mode as done in cuprates
\cite{Fischer-2007a,Levy-2009,Jenkins-2009}. However, still-open
issues render such a quantitative analysis not straightforward in
pnictides. Undoped pnictide compounds present spin-density-wave
metallic order \cite{Luetkens-2008a} in contrast to the
antiferromagnetic insulator cuprate parent compounds,
making of crucial importance to study the
doping-evolution of the energy of collective modes in
different pnictides. In addition, the momentum structure of the
superconducting gap is still subject of debate
\cite{Mazin-2008a,Eremin-2008a,Maier-2009a,Kuroki-2008a}. Other
issue rendering this analysis more complicated for pnictides is
its multiband superconductivity \cite{Ding-2008a} with up to five
FeAs-related bands crossing the Fermi level \cite{Kuroki-2008a}.
Nevertheless, an important question so far not addressed is the
manifestation of collective modes that couple with quasiparticles
in the  excitation spectrum of pnictides.

Here we present the first STM data in
single-crystals of a 1111-pnictide, nearly-optimally-doped
SmFeAsO$_{1 - x}$F$_{x}$ samples with nominal composition $x=0.2$ and
$T_{c} = 45$\,K \cite{Weyeneth-2009a}.
The studied crystals were grown
by means of a high-pressure technique and present
high structural, magnetic and transport quality as quantitatively
reported in Ref. \onlinecite{Karpinski-2009a}
We reveal spectral features that
fingerprint the interaction of quasiparticles with a collective
mode. We report on the pairing-strength-dependence of the mode
energy, of fundamental importance on unveiling the pairing
mechanism. In addition, fits of the low-bias tunneling conductance
lead us to suggest that the momentum structure of the
superconducting gap quite likely presents nodes. We probe the
pnictides' quasiparticle excitation spectrum by mapping the local
tunnel conductance $dI/dV$ versus $V_{bias}$, the tip-sample
voltage difference. The hundred-micron-sized crystals
shown in Fig.\,1 was studied using a home-made low-temperature and
ultra-high-vacuum STM. Reproducible topographic images of as-grown
surfaces reveal growth-terraces with terrace-valley
height-differences of 1 or 2 $c$-axis unit cells, $c=8.47$\,\AA.

Maps of local $dI/dV (V_{bias})$ spectra acquired at 4.2\,K in
nanometer-sized fields-of-view reveal systematic features in the
quasiparticle excitation spectra whose spectral shape and
energy-location spatially varies.
As illustrated in Fig. 1(c),
spectra are asymmetric for occupied and empty sample states.
Three
spectral features are systematically detected for $T < T_{c}$: a
low-bias $dI/dV$ depletion flanked by conductance peaks that at
some locations faint on kinks (black arrows),
a dip in conductance  at $E_{dip}
\sim 10$\,meV (green circles), and a high-bias peak/kink structure
at $E_{2} \sim
20$\,mV (green arrows). Figure 1 (d) shows the low-bias depletion fills in on
warming through $T_{c}$, indicating it is the manifestation of a
superconducting gap $\Delta$. Following common practice
\cite{Fischer-2007a}, we estimate the gap as half the peak-to-peak
energy separation, $\Delta_{p}$. The local $\Delta_{p}$ values
roughly follows a gaussian distribution with a mean $\langle
\Delta_{p} \rangle = 7$\,meV and a full-width at half-maximum of
$\sim 0.14 \langle \Delta_{p} \rangle $, smaller than for cuprates
\cite{Fischer-2007a}. Point-contact spectroscopy studies in single
and polycrystalline samples of SmFeAsO$_{1 - x}$F$_{x}$ found
similar  gap values
\cite{Karpinski-2009a,Chen-2008,Yates-2008,Samuely-2008}, yielding
$2 \Delta/k_{B}T_{c} \sim 3.6$ as in classical superconductors.
This ratio contrasts with roughly two-times larger values found in
STM studies of 122-pnictides \cite{Yin-2009,Massee-2009},
interpreted as an indication of unconventional superconductivity
in these materials.

STM measures the spectral function integrated along the pockets of
the Fermi-surface and therefore lacks direct information on the
k-dependence of the superconducting order parameter.  However,
evidence on the momentum-structure and eventual nodes of $\Delta$
can be obtained by fitting the low-bias $dI/dV$. This issue is
related to the momentum-location of collective modes: a spin resonance
 located at $(\pi,\pi)$, as recently detected in 1111-pnictides
 \cite{Wakimoto-2009a}, implies that the sign of $\Delta$ changes for
  pockets' points connected by a $(\pi,\pi)$ vector
  \cite{Korshunov-2008a,Maier-2009a}. Therefore, we considered
   two pairing symmetries that are compatible with this scenario:
    sign-reversing-$s \pm$ \cite{Maier-2009a} and nodal-$d$. For
    the sign-reversing-$s \pm$ symmetry $\Delta$ changes sign
     between the hole ($\alpha$ bands) and the electron ($\beta$ bands)
     pockets whereas for the nodal-$d$ symmetry $\Delta$ changes
     sign within a given pocket. The fits of  the normalized tunnel
     conductance consider a superconducting and a normal-conducting
     channel amounting to circa 75\% of the total conductance. The
     fits of the superconducting channel are based on a BCS
     quasiparticle excitation spectrum broadened with a finite
     quasiparticles' lifetime parameter and  integrated along a
     single circular hole-pocket centered around the $(0,0)$ point.
     The fitting parameters are the superconducting gap $\Delta$
     and the quasiparticles' lifetime $\tau = \hbar / \Gamma$ with
     $\Gamma$ the Dynes scattering rate. Figure 2 shows the fits of
     average spectra with different $\Delta_{p}$. For both models,
     the fits are robust and of similar quality:  the standard deviation
      of the two fitted parameters is of the order of 5\% for
      $\Delta$ and 10\% for $\Gamma$ for the three average spectra.
      The $\Delta$ and $\Gamma$ values obtained for the two models
      are listed in the caption of Fig. 2.
      The $\Delta$ values obtained
      with the the sign-reversing-$s \pm$ fits yield
       $2 \Delta/k_{B}T_{c} \sim 2.2$, a value too
        small in order to account for the $ T_{c}$ of the samples. In
        addition,the sign-reversing-$s \pm$ fits yield a
        significant
        shortening of quasiparticles' lifetime (see caption of
        Fig.\,2). On the contrary, in the case of the nodal-$d$ symmetry
         the ratio $2 \Delta/k_{B}T_{c} \sim 3$\, is closer to the
         $3.5$ value that
         holds in most classical superconductors. These results
         suggest nodal-$d$ as the most plausible candidate for the
           symmetry of the order parameter in SmFeAsO$_{1 - x}$F$_{x}$.

We now consider the phenomenology of the spectral feature detected
at $E_{2}$. Figure 3 shows two archetypical local $dI/dV$ curves:
a hump-like structure is detected either as a kink or as a peak.
Statistics on individual spectra reveal that $E_{2}$ has a larger spatial
variation (full-width at half-maximum $\sim 40$\,\% of the mean
value) than $\Delta_{p}$, see Fig. 3.
In contrast with the particle-hole
symmetry observed for $\Delta_{p}$, the center of mass of the
$E_{2}$ distributions for empty and occupied sample states are
shifted by 3\,meV.
A similar behavior was reported for a  conductance-shoulder
observed at $\sim 16$\,meV in point-contact
experiments in SmFeAsO$_{1-x}$F$_{x}$ single-crystals from the
 same from the same source than ours \cite{Karpinski-2009a}.
 Recent penetration depth measurements in crystals from the same
 batch suggest a two-gap scenario: one with magnitude in agreement
with our STM data and a smaller of roughly 4\,meV
\cite{Malone-2009a}. This evidence from bulk measurements, and
the locally-observed particle-hole asymmetry of $E_{2}$, casts
doubts on the interpretation of this feature as the fingerprint of
a second superconducting gap.

Alternatively, the hump-like structure can have origin in a
redistribution of spectral weight as a consequence of a
conductance-dip developing at energies $E_{dip}$, see Fig. 3 (a).
In analogy to conventional and cuprate superconductors,
this feature can be interpreted  as a manifestation of the coupling of
quasiparticles with a collective excitation
\cite{Giaever-1962,McMillan-1965,Eschrig-2000,Levy-2009,Jenkins-2009}.
Two possible candidates for the collective excitation are phonon
and spin modes.
Conventional electron-phonon mediated
pairing in pnictides is not likely: theoretical
 calculations for 1111-compounds report the electron-phonon
 coupling is too weak  to explain the materials high $T_{c}$
 \cite{Boeri-2008a}, a conclusion also supported by neutron
  data \cite{Christianson-2008b}.
  On
the other hand, a resonant spin excitation has  been
 detected in the superconducting state of
1111 \cite{Wakimoto-2009a} and 122-pnictides
\cite{Christianson-2008a}. Electron-tunneling data can provide
 information on the energy of the mode that couples
with quasiparticles leading to ascertain which of the two
candidates is relevant for pairing.

Indeed, the quantitative explanation of particular features in the
electron-tunneling spectra of classical superconductors
 stand among the most convincing
validations of the conventional BCS phonon-mediated pairing theory
\cite{Giaever-1962,McMillan-1965}. In the case of cuprates, we
have recently reported \cite{Levy-2009,Jenkins-2009}  that: 1) the
dip feature is the fingerprint of quasiparticles coupling with an
energy and momentum-localized collective excitation,  2) the
energy of the collective mode can be directly read out from
$dI/dV$ curves as $\Omega = |E_{dip} - \Delta_{p} |$, with 1\,meV
uncertainty. Here we will follow the same approach: profiting from
the spatial inhomogeneity of $\Delta_{p}$  we study the evolution
of $\Omega$ with the pairing strength. Figure 4 shows that the dip
shifts towards the coherence peak when $\Delta_{p}$ increases,
implying that the energy of the collective mode that couples with
quasiparticles is anticorrelated with the local pairing strength.
This result bears remarkable similarity with the phenomenology
reported in the case of cuprates \cite{Eschrig-2006}.

The value of the local collective mode energy shown in Fig.\, 4
ranges between 8 and 2\,meV.
The typical energy of the spin
resonance detected by  neutrons in
LaFeAsO$_{1-x}$F$_{x}$, the few available data in 1111-pnictides,
is of the order of 11\,meV \cite{Wakimoto-2009a}.
 This value is
quite sensitive to the compound and the doping level,
making of great importance to obtain neutron data in the
particular material studied here. Nevertheless, an important
result of Ref.\cite{Wakimoto-2009a} is that the spin resonance is
only detected for superconducting samples. On the contrary, phonon
modes are detected in parent and doped compounds, with their
energy  not significantly changing with temperature
\cite{Christianson-2008b}. These findings suggest that the
collective mode revealed through our analysis of STM data is quite
likely the ubiquitous $(\pi,\pi)$ spin resonant mode, signposting
towards spin-fluctuation-mediated pairing in pnictides.

\newpage

\begin{figure}[hhh]
\includegraphics[width=0.7\columnwidth,angle=0]{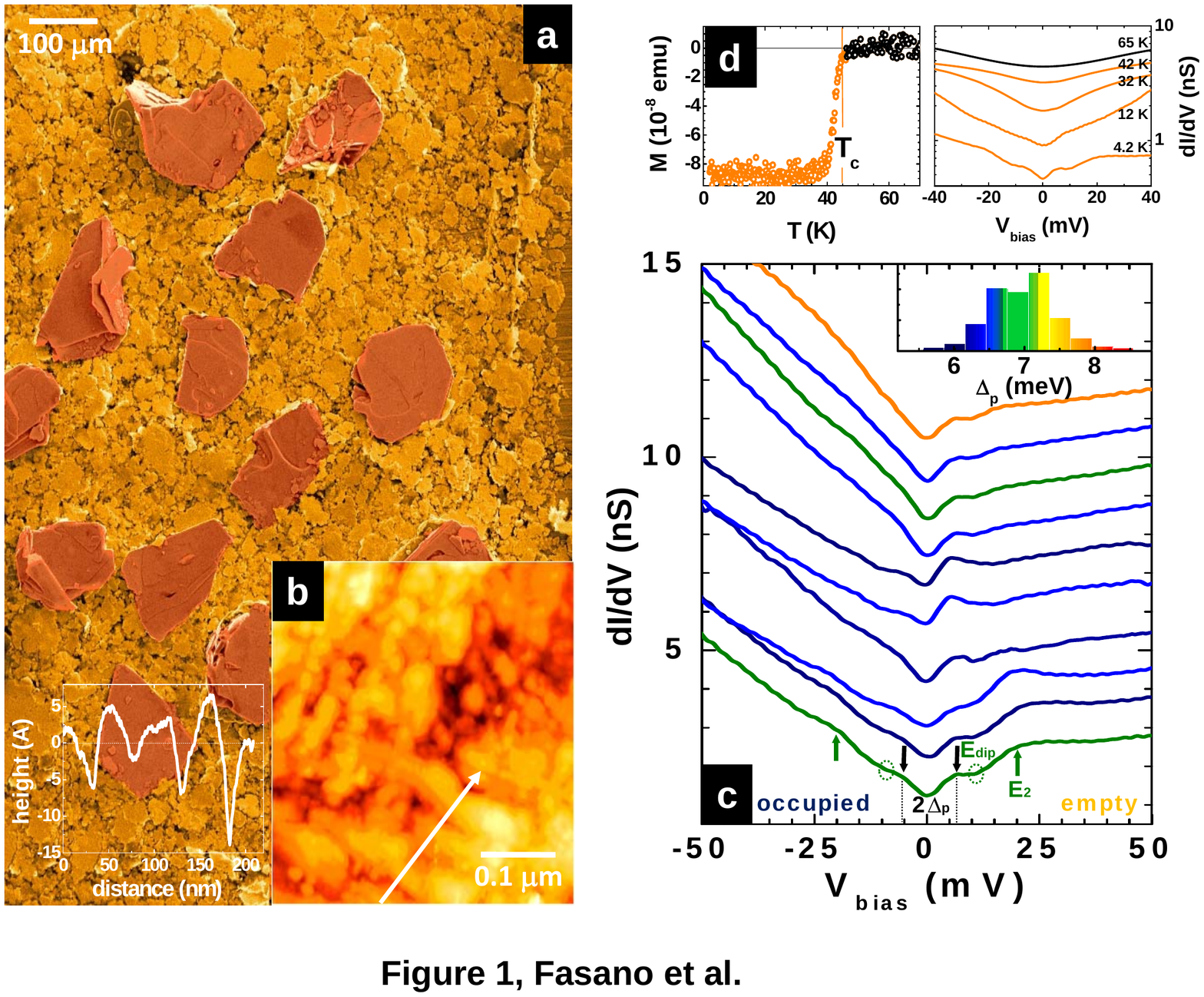}
\caption[]{ (a) SEM image of the
 measured single-crystals from the same batch.
(b) Right panel: STM topographic image of an as-grown surface
depicting growth-terraces with a rms roughness smaller than
0.1\,nm. Left panel: Height profile along the path indicated with
an arrow: terrace-valley height differences are either one or two
crystal unit-cell parameters ($c=8.47$\,\AA ). (c) Local $dI/dV$
spectra (vertically shifted) acquired at 4.2\,K along an 8\,nm
path. Spectra are color-coded according to its experimental gap
value calculated as half the peak-to-peak/kink (black arrows)
energy separation. The color-convention is that of the local
$\Delta_{p}$ histogram of the insert resulting from 4500 local
spectra mapped in a $40 \times 40$\,nm$^2$ field-of-view. Green
arrows indicate the high-bias kink  located at $E_{2}$ and green
circles the local dip at $E_{dip}$. (d) Left panel:
Zero-field-cooled magnetic moment of a crystal yielding
$T_{c}=45$\,K and a transition width of 5\,K
\cite{Weyeneth-2009a}. Right panel: temperature-evolution of
average $dI/dV$ curves (in a $40 \times 40$\,nm$^2$
field-of-view). In the STM measurements an
electrochemically-etched Ir tip served as the ground electrode and
thus negative (positive) bias refers to occupied (empty) sample
states. The tunnel-junction regulation parameters were of 80\,meV
and 1\,nA. Spectra were acquired by means of a lock-in technique
\cite{Fischer-2007a} with a bias modulation of 1.5\,mV$_{rms}$ and
an energy resolution of 0.6\,mV.}
\end{figure}

\begin{figure}
\includegraphics[width=0.7\columnwidth, angle=0]{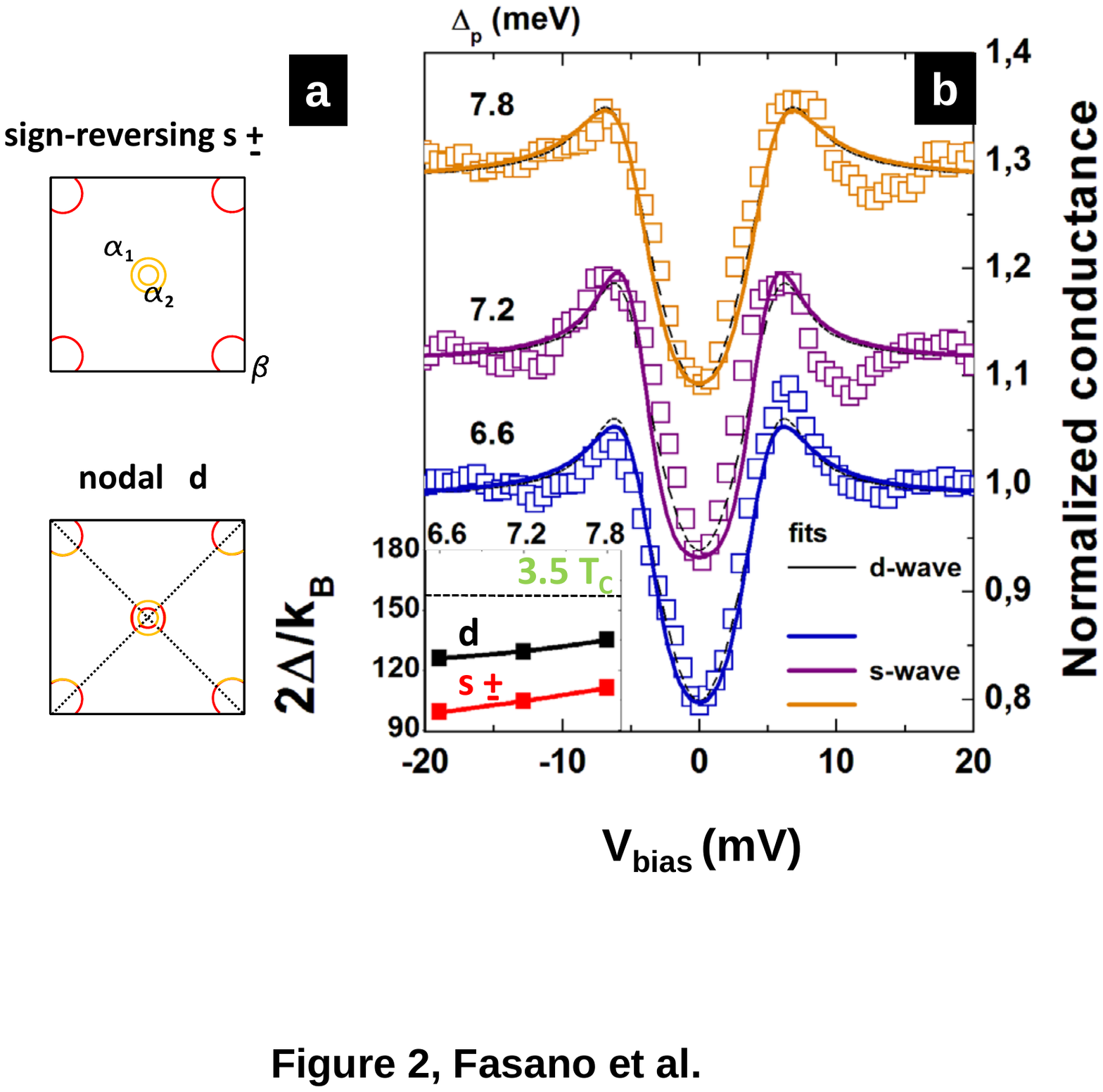}
\caption[]{ (a) Schematics of the 1111-pnictides Fermi surface
pockets ($\alpha$-hole and $\beta$-electron sheets) including
information on possible scenarios for the momentum-structure of
the gap. The orange (red) locations represent a positive
(negative) gap. Upper panel: sign-reversing $s \pm$ symmetry.
Lower pannel: nodal $d$ symmetry with the dotted lines depicting
the location of the gap nodes. (b) Average normalized spectra at
4.2\,K obtained from hundreds of local  $dI/dV$ curves with same
$\Delta_{p}$ (open symbols). Full lines are fits considering the
sign-reversing $s \pm$ (orange, purple,blue) and nodal $d$ (black)
symmetries. A 4.2\,K thermal broadening and a finite
quasiparticles lifetime $\tau = \hbar / \Gamma$, with $\Gamma$ the
scattering rate, were considered. Sign-reversing $s \pm$ fits
yield $\Delta$  4.2, 4.3 and 4.8\,meV for the average spectra with
increasing $\Delta_{p}$. In the case of nodal-$d$ fits the fitted
$\Delta$ values are of
 5.1, 5.2, and 5.8\,meV,
respectively. On increasing $\Delta_{p}$ the scattering rate
obtained in the sign-reversing-$s \pm$ fits are 0.9, 0.7 and 1.5
\,meV
 roughly twice the
values obtained in the nodal-$d$ fits, 0.3, 0.3, and 0.9\,meV,
respectively. Insert: Evolution of $2 \Delta/k_{B}$ with
$\Delta_{p}$ for the two models. The dotted line indicates the
weak-coupling BCS limit of $3.5 T_{c}$ holding for  classical
superconductors.}
\end{figure}

\begin{figure}
\hspace{-3.4cm}
\includegraphics[width=0.7\columnwidth, angle=0]{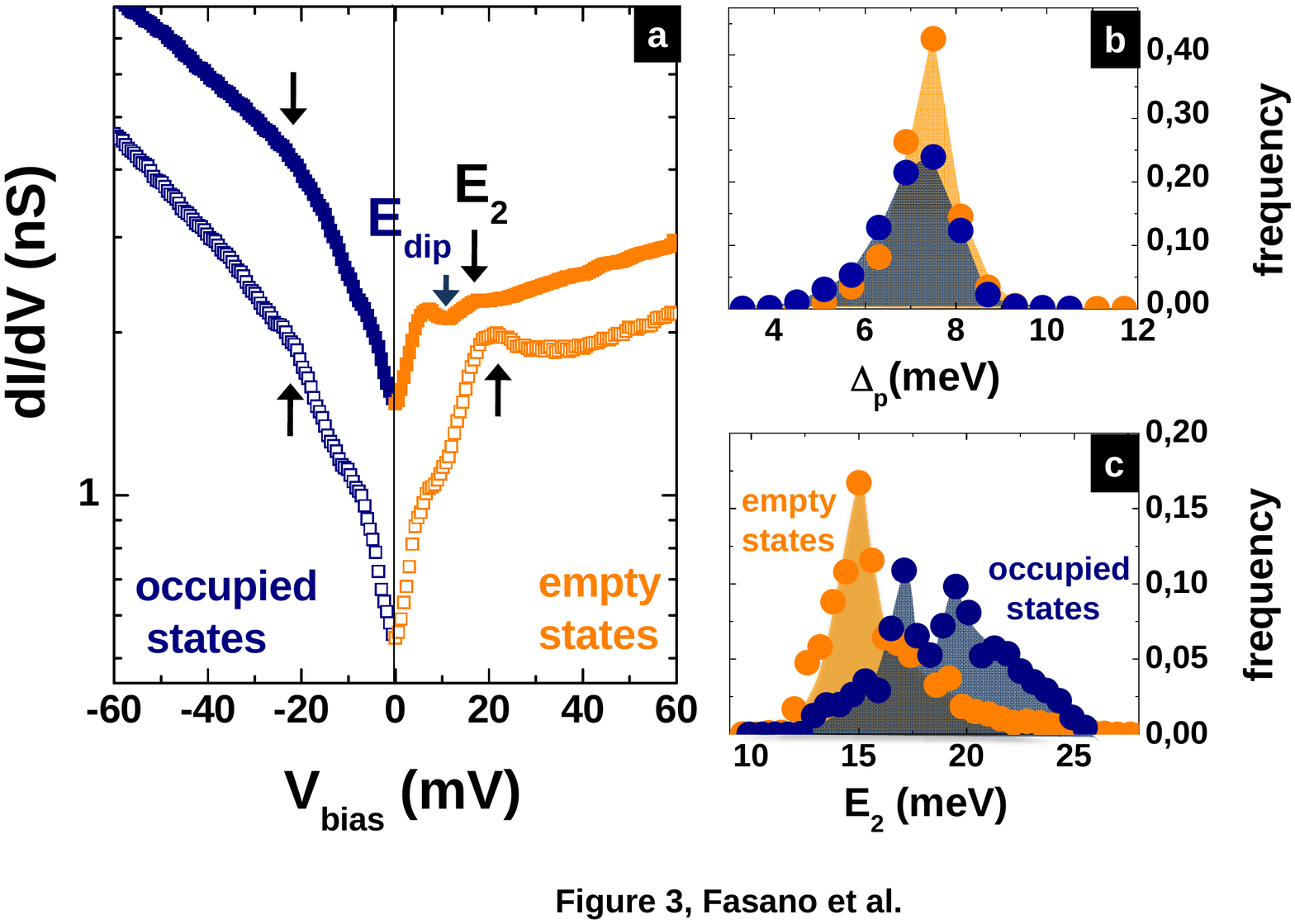}
\caption[]{(a) Two archetypical local $dI/dV$ curves (vertically
shifted) illustrating the spectral shape of the high-energy
feature detected at $E_{2}$ (black arrows) and of the dip-like
feature observed at $E_{dip}$ (blue arrow). (b) Histograms of the
energy-location of the coherence peaks/kinks, $\Delta_{p}$, for
empty (orange) and occupied
  (blue) sample states. (c) Histogram of the
 local $E_{2}$ values: the distribution for
 the empty and occupied-sample states  have mean values displaced
  in $\sim 3$\,meV.
 Histograms obtained  from 4500 spectra acquired over a
 $40 \times 40$\,nm$^2$ field-of-view. }
\end{figure}

\begin{figure}
\hspace{-4.5cm}
\includegraphics[width=0.7\columnwidth, angle=0]{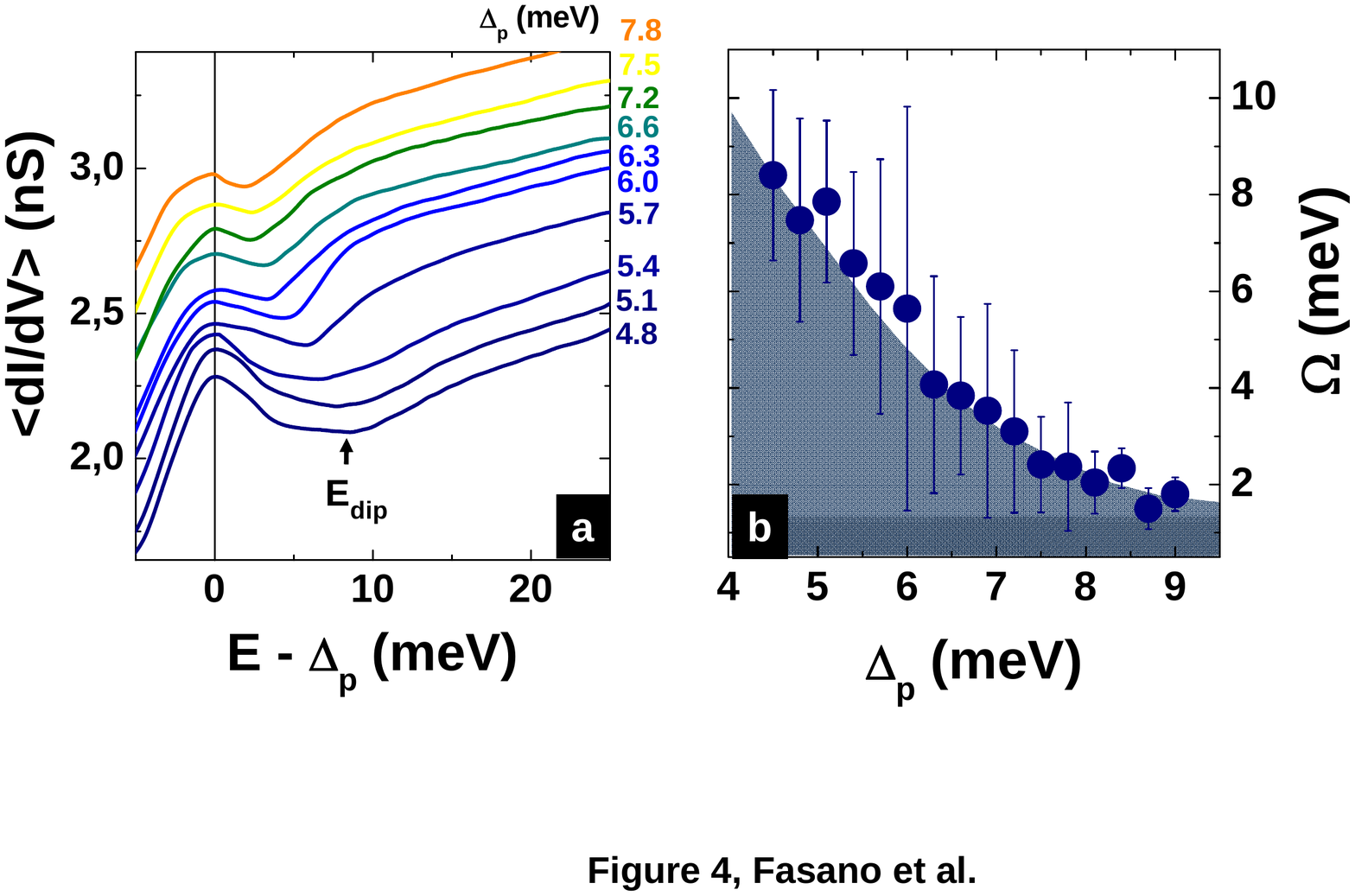}
\caption[]{ (a) Average $dI/dV$ curves for local spectra with the
same $\Delta_{p}$ plotted as a function of the sample bias minus
$\Delta_{p}$ (for the empty sample states). On increasing
$\Delta_{p}$ the dip located at $E_{dip}$ shifts towards the
coherence peak. (b) $\Delta_{p}$-dependence of the collective-mode
energy  measured as the peak-to-dip energy separation, $\Omega=
E_{dip} - \Delta_{p}$ .  The darker region indicates the 1\,meV
uncertainty in determining $\Omega$. }
\end{figure}

\end{document}